\documentclass[conference]{IEEEtran}
\usepackage{cite}
\usepackage{amsmath,amssymb,amsfonts}
\usepackage{graphicx}
\usepackage{textcomp}
\usepackage{xcolor}
\usepackage{multirow}
\usepackage{array}
\usepackage{makecell}
\usepackage{enumerate}
\usepackage{tikz}
\usepackage{algpseudocode}
\usepackage{comment}
\usepackage[english]{babel}
\newtheorem{thm}{Theorem}

\newtheorem{cor}{Corollary}

\newtheorem{defn}{Definition}

\newtheorem{exmp}{Example}

\newtheorem{rem}{Remark}

\def\BibTeX{{\rm B\kern-.05em{\sc i\kern-.025em b}\kern-.08em
    T\kern-.1667em\lower.7ex\hbox{E}\kern-.125emX}}
\begin{document}

\title{Secretive Coded Caching from PDAs\\
}

\author{\IEEEauthorblockN{ Shreya Shrestha Meel}
\IEEEauthorblockA{\textit{Department of Electrical Communication Engineering}\\
\textit{Indian Institute of Science,}\\
Bengaluru 560012, KA, India \\
shreyameel@iisc.ac.in}
\and
\IEEEauthorblockN{ B. Sundar Rajan}
\IEEEauthorblockA{\textit{Department of Electrical Communication Engineering} \\
\textit{Indian Institute of Science,}\\
Bengaluru 560012, KA, India \\
bsrajan@iisc.ac.in}
}

\maketitle

	\begin{abstract}
		The coded caching problem with secrecy constraint i.e., the users should not be able to gain any information about the content of the files that they did not demand, is known as the secretive coded caching problem. This was proposed by Ravindrakumar et al. in the paper titled ``Private Coded Caching'' that appeared in \emph{ IEEE Transactions on Information Forensics and Security}, 2018 and is characterised by subpacketization levels growing exponentially with the number of users. In the context of coded caching without secrecy, coded caching schemes at subexponential subpacketization levels are feasible by representing the caching system in the form of a Placement Delivery Array (PDA) and designing placement and delivery policies from it. Motivated by this, we propose a secretive coded caching scheme with low subpacketization using PDA, for users with dedicated caches in the centralized setting. When our scheme is applied to a special class of PDA known as MN PDA, the scheme proposed by Ravindrakumar et al. is recovered. 
	\end{abstract}


\section{Introduction}
\label{sec1}	
Caching is an effective strategy to balance the disparity between peak and off-peak traffic load in content delivery networks by making parts of content available to the users in their local caches. The seminal work \cite{MaNcentCc} established that, during peak hours, congestion of the shared link between the server and users can be further reduced by transmitting coded messages, that are useful to more than one user simultaneously. This approach is known as \emph{coded caching}. The authors consider a network with $K$ users, each equipped with a local cache large enough to store $M$ files. The users are connected to a server through a shared, error-free link. The server hosts a library of $N$ files of equal size. The system operates in two phases-\emph{placement phase} and \emph{delivery phase}. The placement phase concerns with the server filling the users' caches with some fraction of \emph{all the files}, without knowing the future demands of the users. In the delivery phase, users reveal their file requests in response of which, the server sends coded multicast transmissions over the broadcast link such that, the users can, from the transmissions and cache contents, recover their requested file. The goal is to minimise the \emph{rate} i.e., the number of bits transmitted over the shared link, normalised by file size. 
The caching policy in \cite{MaNcentCc} gives a user illegitimate access to parts of files it has not requested. This motivates the formulation of the \emph{secretive coded caching} problem, introduced in \cite{SecretiveCc}. In this setting, the users should not gain, either from the cache contents or server transmissions, any information about the content of the files that they did not request. The delivery rate achieved by a secretive coded caching scheme with memory $M$ is known as \emph{secretively achievable rate} and we denote it by $R^s(M)$. The optimum secretively achievable rate $R^{s\star}(M)$ is given by $\inf\{R^s:R^s(M) \text{ is secretively achievable}\}$.
\par Secretive coded caching problem was first studied in \cite{SecretiveCc,PrivateCc} where the authors proposed an achievable coded caching scheme imposing the constraint of \textit{perfect secrecy}, under both centralized and decentralized settings. Their general achievable scheme also guaranteed secure delivery against eavesdroppers as in\cite{SecureDelivCc}, in addition to satisfying the secrecy constraint. They also gave a lower bound on the optimum secretively achievable rate $R^{s\star}$ based on the cut-set bound, and showed that the achievable rate was at a constant multiplicative gap of $16$ from the lower bound, for certain memory regimes. This problem was extended to Device-to-Device caching networks in \cite{D2DSecureCc} and to Combination networks in \cite{SecCombi}. The authors in \cite{SecretiveCollud} considered secretive coded caching in a setting where at most $l$ out of the $K$ users are untrustworthy and might collude among themselves. They provided an upper and a lower bound on the optimal secretively achievable rate. On setting the value of $l$ to unity, their scheme reduces to the original secretive coded caching scheme \cite{PrivateCc}. In \cite{ImprovedSecretiveCc}, the authors proposed a new average secretively achievable scheme for uniformly distributed demands, where, unlike the scheme in \cite{PrivateCc}, the server transmitted some messages without encrypting them with keys. By exploiting the redundancy of demands, they leveraged the ideas presented in \cite{YMACc} to reduce the number of transmissions since some messages could be obtained simply from a linear combination of others. They identified the messages which should be sent without encryption so that the non-transmitted messages can be recovered from their linear combinations while maintaining the secrecy. We now briefly review the scheme in \cite{PrivateCc}.
\subsection{The Ravindrakumar et al. Secretive Coded Caching Scheme}
\label{sec1A}
The general scheme for centralized setting, in \cite{PrivateCc} involves encoding each file in the library using a $\big(\binom{K-1}{t-1},\binom{K}{t}\big)$-\emph{non-perfect secret-sharing scheme} where $t=\frac{K(M-1)}{N+M-1}\in\{0,1,\ldots,K-2\}$ and with a $(K-1,K)$ secret-sharing scheme when $M=N(K-1)$. The placement phase operates over private links between server and users, by placing these \emph{shares} along with randomly chosen \emph{unique keys}, each of length equal to that of a share. When demands are revealed in the delivery phase, the server sends $\binom{K}{t+1}$ multicast messages over the shared link. Each message is an XOR'ed  combination of shares, further XOR'ed with a unique key. The secretively achievable rate $R^s(M)$ is given by:
\begin{align}
\label{eq1}
R^s(M)=
\begin{cases}
\frac{K(N+M-1)}{N+(K+1)(M-1)} & \text{ for }  M=\frac{Nt}{K-t}+1\\
1, & \text{ for }  M=N(K-1),
\end{cases}
\end{align}
where $t\in\{0,1,\ldots,K-2\}$. Further, the lower convex envelope of these points gives an achievable rate, for all $M\in[1,N(K-1)]$ by memory sharing.
\par \emph{Subpacketization problem:}
	The number of sub-files into which a file is split into is known as \emph{subpacketization level}. To achieve the rate in \eqref{eq1}, the required subpacketization level is $\binom{K}{t}-\binom{K-1}{t-1}=\binom{K-1}{t}$.
	  Essentially, this secretive coded caching scheme is based on the  scheme in \cite{MaNcentCc}, and hence is characterized by subpacketization levels growing exponentially with $K-1$. For large networks with many users, high subpacketization level implies that, the size of the header and trailer bits which are added during server transmission dominates over the size of a message. This increases the delivery rate and defeats the purpose of coded caching. Moreover, to avoid this, the file sizes need to be in terabytes, which limits the practical applicability of the scheme. It was shown in \cite{PDAmain} that with mathematical structures, known as Placement Delivery Arrays (PDAs), the placement and delivery phases for coded caching schemes can be derived at low subpacketization levels that grow sub-exponentially with the number of users, and hence, are more practical. 
\par In this work, we study how given any Placement Delivery Array(PDA), a secretive coded caching scheme can be obtained. The proposed scheme also guarantees secure delivery against an eavesdropper who might wiretap on the shared link. We show that, when our scheme is applied to the Maddah-Ali Niesen (MN) PDA \cite{PDAmain}, it coincides with the general secretively achievable scheme in \cite{PrivateCc}. To implement secretive coded caching in a centralized setting, larger cache memory ($M\geq1$) and higher achievable rate ($R^{s}(M)\geq 1$) are paid as a penalty for secrecy. However, the required number of sub-files into which each file is partitioned into, is less by an additive gap. In this light, we point out that any secretive coded caching scheme is characterized by lower subpacketization level compared to that without secrecy.

\subsection{Contributions}\label{sec1B}
\par The contributions of this paper can be summarized as:
	\begin{itemize}
		\item  Given a $(K,F,Z,S)$ Placement Delivery Array (PDA) (see Sec. \ref{sec2A}) and a $(K,M,N)$ caching system with dedicated caches satisfying $\frac{M-1}{N+M-1}=\frac{Z}{F}$, we show that a secretive coded caching scheme is obtained with achievable rate $\frac{S}{F-Z}$.
		\item The subpacketization level required for our scheme is $F-Z$, while it is $F$ for the non-secretive scheme as in \cite{PDAmain}.
		\item Our scheme can be applied to any PDA to produce a secretively achievable scheme, and when implemented on a special class of PDA known as Maddah-Ali Niesen (MN) PDA, it coincides with the scheme in \cite{PrivateCc}. 
	\end{itemize}
\par The rest of the paper is organized as follows. Section \ref{sec2} presents some preliminaries for our study. In Section \ref{sec3}, we describe the system model. In Section \ref{sec4}, we discuss the main result of the paper and in Section \ref{sec5}, we describe the proposed coded caching scheme. We provide a comparison in performance of our scheme with respect to that in \cite{PrivateCc} and \cite{PDAmain} in Section \ref{sec6}. Finally, Section \ref{sec7} concludes the paper.
\section{Background and Preliminaries}
\label{sec2}
In this section, we briefly review the ideas of PDA, secret-sharing schemes and Cauchy matrix to demonstrate this work.
\subsection{Placement Delivery Array (PDA)}
\label{sec2A}
\begin{defn}[PDA\cite{PDAmain}]
	\label{defPDA}
	For positive integers $K$, $F$, $Z$ and $S$, an $F\times K$ array $\mathbf{P}=(p_{j,k}), j\in[F], k\in[K]$, composed of a specific symbol `` $*$ '' and $S$ positive integers from $[S]$, is called a $(K,F,Z,S)$ PDA if it satisfies the following three conditions:
	\begin{enumerate}[C1.]
		\item The symbol `` $*$ '' appears $Z$ times in each column.
		\item Each integer occurs at least once in the array.
		\item For any two distinct entries $p_{j_1,k_1}$ and $p_{j_2,k_2}$, $p_{j_1,k_1}=p_{j_2,k_2}=s$ is an integer only if:
		\begin{enumerate}[a.]
			\item $j_1\neq j_2$, $k_1\neq k_2$ i.e., they lie in distinct rows and columns, and
			\item $p_{j_1,k_2}=p_{j_2,k_1}=*$, i.e., the corresponding $2\times 2$ sub-array formed by the rows $j_1$, $j_2$ and columns $k_1$, $k_2$ must be of the form
			\begin{align*}
			\begin{bmatrix}
			s & *\\ 
			* & s
			\end{bmatrix} \hspace{5 mm}\mbox{or}\hspace{5 mm}
			\begin{bmatrix}
			* & s\\
			s & *
			\end{bmatrix}.
			\end{align*}
		\end{enumerate}
	\end{enumerate}
	
\end{defn}
\begin{exmp}
\label{ex1}
The array $\mathbf{P}$  is a $(6,4,2,4)$ PDA.
	\begin{align}
	\label{pdaex}
	\mathbf{P}=
	\begin{bmatrix}
	* & 2 & * & 3 & * & 1\\
	1 & * & * & 4 & 2 & *\\
	* & 4 & 1 & * & 3 & *\\
	3 & * & 2 & * & * & 4\\
	\end{bmatrix}
	\end{align}
\end{exmp}
\par By Theorem 1 of \cite{PDAmain}, based on a $(K,F,Z,S)$-PDA $\mathbf{P}=(p_{j,k}), j\in[F], k\in[K]$, a non-secretive coded caching scheme with subpacketization level $F$ for a $(K,M,N)$ cache network satisfying $\frac{Z}{F}=\frac{M}{N}$ can be obtained. Every PDA represents a coded caching scheme. For non-secretive coded caching, there are as many rows in a PDA as the subpacketization level, and each row, indexed by $[F]$ represents a sub-file, while each column, indexed by $[K]$ represents the corresponding user. The placement and delivery policies are as follows:
	\par\emph{Placement phase:} The same placement policy is followed for every file $W^n,n\in[N]$. First, each file of $B$ bits is partitioned into $F$ disjoint sub-files each of size $B/F$ bits such that $W^n=\{W^n_1,W^n_2\ldots,W^n_F\}$. The content placed in the cache of the $k^{th}$ user is given by:
	\begin{align}
	    \label{plcmnt}
	    Z_k=\big\{W^n_j:p_{j,k}=*,\forall j\in[F],\forall n\in [N]\big\}.
	\end{align}
	By C1. of Definition \ref{defPDA}, each user stores $N.Z$ sub-files each of $B/F$ bits. The normalized size of the cache $M=\frac{1}{B}(N.Z.B/F)=\frac{NZ}{F}$ is satisfied.
	\par \emph{Delivery phase:} Let user $k\in [K]$ demand the file $W^{d_k},d_k\in [N]$ from the library. Every distinct integer in the PDA corresponds to a multicast message, leading to a total of $S$ transmissions. Therefore, the server transmits messages in $S$ time slots.  In time slot $s\in [S]$, it transmits:
	\begin{align}\label{tx1}
	   \bigoplus_{p_{j,k}=s,j\in[F],k\in[K]}W^{d_k}_j 
	\end{align}
	where $\bigoplus$ denotes bit-wise XOR operation. On receiving the $S$ messages, the requested files can be retrieved by the users. This is because, from condition C3. of Definition \ref{defPDA}, in each message, except the sub-file of the file demanded by a user, the remaining sub-files are available in the user's cache and can be eliminated (XOR'ed out) \cite{PDAmain}. Thus, the rate achieved is $R(M)=R(\frac{NZ}{F})=\frac{S}{F}$ at a subpacketization of $F$.
\begin{defn}[$g$-regular PDA\cite{PDAmain}]
	An array $\mathbf{P}$ is said to be a $g$-regular $(K,F,Z,S)$ PDA if it satisfies C1, C3 of Definition \ref{defPDA} and the following condition:
	\begin{enumerate}[C2'.]
		\item Each integer appears $g$ times in $\mathbf{P}$ where $g$ is a constant.
	\end{enumerate}
\end{defn}
The PDA in Example \ref{ex1} is a  $3-(6,4,2,4)$ PDA since every integer occurs exactly $3$ times in the array. The significance of $g$ is that, every multicast message in \eqref{tx1} is an XOR'ed combination of exactly $g$ sub-files.

\par{\emph{Maddah-Ali Niesen PDA (MN PDA)\cite{PDAmain}:}}
The equivalent PDA representation of the Maddah-Ali Niesen centralized coded caching scheme \cite{MaNcentCc} is known in the literature as the MN PDA. Here, for each integer $t=\frac{KM}{N}\in\{0,1,...,K\}$, we have a corresponding MN PDA, where $t$ indicates the number of ``$*$''s present in each row of the PDA. $K$ is the number of columns and $F=\binom{K}{t}$, is the number of rows of the PDA. Consider the set, $\mathcal{S}\subseteq[K],|\mathcal{S}|=t+1$. We now arrange the $t+1$ sized subsets $\mathcal{S}\in \binom{[K]}{t+1}$ of $[K]$ in lexicographic order to obtain an ordered list of $\binom{K}{t+1}$ elements. Let $\phi(\mathcal{S})$ reflect the order of $\mathcal{S}$ in this list.  Then, we get a one-one correspondence $\phi$ as:
\begin{align}\phi:\binom{[K]}{t+1}\mapsto \Bigg[\binom{K}{t+1}\Bigg]\end{align}
For example, if $K=4$ and $t=2$, then $\phi(\{1,2,3\})=1$, $\phi(\{1,2,4\})=2$,  $\phi(\{1,3,4\})=3$ and  $\phi(\{2,3,4\})=4$.
Each row of of the PDA is indexed by $\mathcal{T}$, where $\mathcal{T}\subset [K], |\mathcal{T}|=t$ and columns are indexed from $1,2,\ldots, K$. Then, the MN PDA is defined as a $\binom{K}{t}\times K$ array $\mathbf{P'}=(p'_{\mathcal{T},k}),\mathcal{T}\in \binom{[K]}{t}, k\in [K]$ as:
\begin{align*}
p'_{\mathcal{T},k}=
\left\{
\begin{array}{ll}
\phi(\mathcal{T}\cup \{k\}) & \mbox{, if }k\notin\mathcal{T} \\
\mbox{*} &\mbox{, otherwise}
\end{array}
\right.
\end{align*}
The MN PDA $\mathbf{P}$ is a regular PDA with $g=t+1$ and $(K,F,Z,S)$ given by $(K,\binom{K}{t},\binom{K-1}{t-1},\binom{K}{t+1})$ respectively, where $\frac{t}{K}=\frac{Z}{F}=\frac{M}{N}$.
\subsection{Secret-Sharing schemes}
\label{sec2B}
The secretive coded caching algorithms in \cite{SecretiveCc, PrivateCc,D2DSecureCc,
	SecCombi,SecretiveCollud,ImprovedSecretiveCc} rely on $(m,n)$-\emph{non-perfect} secret sharing schemes, where $m,n>0, \ m<n$. The values of $m$ and $n$ are determined by the caching system requirements. 
\begin{defn}[(m,n)-non-perfect secret sharing scheme\cite{Ogata}]
	 Let $\mathbb{F}_q$ be a finite field where $q$ is a prime or prime power. On a set of participants, $P =\{P_0,P_1,P_2,\ldots,P_n\}$, a distinguished participant $P_0$, known as the \emph{dealer} selects a \emph{secret} $s_0$ randomly from $\mathbb{F}_q^{n-m}$. Let $S_0$ be the random variable induced by it, and $H(S_0)>0$. The participants from $P_1,\ldots, P_n$, known as \emph{players}, are ignorant of this secret. The dealer distributes \emph{shares} of $s_0$ among the players. Let player $P_i$ be given the share $s_i$, which is distributed over $\mathbb{F}_q$ and let $S_i$ be the random variable induced by it. The random vector of shares having realization $s$, is given by $S=(S_i)_{i=1}^{n}\in \mathbb{F}_q^n$. Given a distribution of secrets represented by $S_0$ and a distribution of random bit-strings over $\mathbb{F}_q^m$ represented by the random variable $R$ having realization $r$, suppose that there is a mapping,
	\begin{align*}
	\Pi:\mathbb{F}_q^{n-m}\times\mathbb{F}_q^m\mapsto \mathbb{F}_q^n.
	\end{align*} 
	with $(s_1,s_2,\ldots, s_n)=\Pi(s_0,r)$ such that $H(S_0|S_{i_1},S_{i_2},\ldots,S_{i_m})=H(S_0)$ where $i_1,i_2,\ldots,i_m\subset[n]$ and $H(S_0|S)=0$, i.e., if $m$ or less players pool their shares, it reveals no information about the secret and only if all the $n$ players pool their shares shares $s=(s_1,s_2,\ldots,s_n)$, it is sufficient to exactly reconstruct the secret. Also, pooling any number of shares between $m+1$ to $n-1$ reveals partial information about $s_0$. Then, $(\Pi,S_0,S)$ is said to be an $(m,n)$-non-perfect secret sharing scheme. 
\end{defn}
\par Essentially, a non-perfect secret sharing scheme is one that allows certain collections of shares to gain some information about the secret. On the contrary, an $(m,n)$-perfect secret sharing scheme, such as Shamir's \emph{threshold scheme} \cite{Shamir},\cite{SecureMPC} allows a collection of shares to either obtain all or no information about the secret. Here, access to any $m+1$ out of $n$ shares completely reveals the secret, whereas access to any $m$ or less shares, reveals no information about the secret. For the non-perfect schemes, the size of each share is $\frac{1}{n-m}$ times the size of the secret\cite{Farras},\cite{Ogata}, whereas in the perfect schemes, it equals the size of the secret. Hence, non-perfect secret sharing schemes are advantageous in terms of saving resources (memory, transmission load). The \emph{ramp threshold} secret sharing schemes as in \cite{BlakelyMeadows,Yamamoto} are examples of non-perfect secret sharing schemes. 
\subsection{Cauchy matrix}
\begin{defn}[Cauchy matrix\cite{Cauchy}]\label{cauchy}
	Let $\mathbb{F}_q$ be a finite field, with $q\geq u+v$. Let $X=\{x_1,x_2,\ldots,x_u\}$ and $Y=\{y_1,y_2,\ldots,y_v\}$ be two distinct sets of elements in $\mathbb{F}_q$ with $X\cap Y=\phi$ such that:
	\begin{enumerate}[i.]
		\item $\forall i\in[u],\forall j\in[v]$ $:x_i-y_j\neq 0$.
		\item $\forall i,j \in [u]: i\neq j$, $x_i\neq x_j$ and $\forall i,j\in [v], i\neq j$, $y_i\neq y_j$.
	\end{enumerate}
	where all operations are performed over $\mathbb{F}_q$. Then, the matrix,
	\begin{align}
	\begin{bmatrix}
	\frac{1}{x_1-y_1} & \frac{1}{x_1-y_2} & \ldots & \frac{1}{x_1-y_v}\\
	\frac{1}{x_2-y_1} & \frac{1}{x_2-y_2} & \ldots & \frac{1}{x_2-y_v}\\
	\vdots &\vdots & \ddots &\vdots\\
	\frac{1}{x_{u-1}-y_1} & \frac{1}{x_{u-1}-y_2} & \ldots & \frac{1}{x_{u-1}-y_v}\\
	\frac{1}{x_u-y_1} & \frac{1}{x_u-y_2} & \ldots & \frac{1}{x_u-y_v}\\
	\end{bmatrix}
	\end{align} is a Cauchy matrix of dimension $u\times v$. A Cauchy matrix has full rank, and each of its sub-matrices is also a Cauchy matrix. 
\end{defn}

\section{System Model}
\label{sec3}
We consider the dedicated caching system with centralized placement \cite{MaNcentCc}. There are $N$ files in the library, labelled as the set $W^{[N]}:=\{W^1,W^2,\ldots, W^N\}$. Each file is of $B$ bits and can be viewed as a random variable generated independently from the set $\{0,1\}^B$. There are $K$ users, each equipped with a cache that can store up to $MB$ bits, i.e., $M$ files, where $M\in[1,N(K-1)]$. It is assumed that the placement phase takes place over private links between the server and users. The content placed at user $k$'s cache is denoted by $Z_k$. The delivery phase takes place over the shared, error-free link between the server and users. During the delivery phase, $\forall k \in [K]$, user $k$ requests for a file $W^{d_k}, \ d_k\in[N]$ independent of other users. The demand vector given by $\mathbf{d}=(d_1,d_2,\ldots,d_K)$ is conveyed to the server. The server broadcasts the message vector $X_\mathbf{d}$ of $R^s(M)B$ bits such that each user $k\in[K]$, by utilising $Z_k$ and $X_{\mathbf{d}}$, is able to exactly recover the file it wants. That is, 
\begin{align}
\label{corr}
H(W^{d_k}\lvert Z_k,X_{\mathbf{d}})=0 \hspace{2mm} \forall \mathbf{d}\in[N]^K, \forall k\in [K].   
\end{align}
\par The quantity $R^s(M)$ is the secretively achievable rate for cache memory $M$ and $(M,R^s)$ is a secretively achievable memory-rate pair. To quantify the notion of secrecy, the authors in \cite{PrivateCc} defined \emph{information leakage} $L$ as:
\begin{align}
L=\max_{k\in[K]}\max_{\mathbf{d}\in[N]^K}I(W^{[N]\setminus d_k};Z_k,X_{\mathbf{d}}).
\end{align}
	where $I(;)$ is the mutual information and $W^{[N]\setminus d_k}$ is the set of all files except the one requested by user $k$. Secretive coded caching problem requires fulfilling \emph{perfect secrecy}, that is $L=0$,
	\begin{align}
	\label{sec}
	    I(W^{[N]\setminus d_k};Z_k,X_{\mathbf{d}})=0, \ \forall \mathbf{d}\in[N]^K, \ \forall k\in [K].
	\end{align}
 A secretive coded caching scheme has to satisfy \eqref{sec}, in addition to \eqref{corr}. 
\section{Main Result}\label{sec4}
In this section, we state the main result of the paper.
\begin{thm}
	\label{scc_rate}
	For any given $(K,F,Z,S)$-PDA, the secretively achievable rate corresponding to the caching scheme with $M=\frac{NZ}{F-Z}+1$, is given by:
	\begin{align}\label{eq_pda_rate}
	R^s(M)=\frac{S}{F-Z}.
	\end{align}
\end{thm}
 \begin{IEEEproof}
 The proof of this result is based on the proposed scheme in Section \ref{sec5A} and the rate calculation in Section \ref{pda_rate}.
\end{IEEEproof}
\begin{cor}\label{cor1}
	If the PDA is $g$-regular, then the secretively achievable rate in \eqref{eq_pda_rate} can be expressed as:
	\begin{align}\label{regpda_rate}
	R^s(M)=\frac{K}{g}.
	\end{align}
\end{cor}
\begin{IEEEproof}
	To prove this, we count the number of integers in the PDA in two different ways. On one hand, since each column has $F-Z$ integers, there are a total of $K(F-Z)$ integers in the array. On the other hand, each integer occurs exactly $g$ times, hence, the total number of integers in the array is $Sg$. Therefore, using this fact and Theorem \ref{scc_rate}, we have:
	\begin{align*}
	Sg=K(F-Z)\implies \frac{S}{F-Z}=\frac{K}{g}.
	\end{align*}
\end{IEEEproof}
\begin{cor}\label{cor2}
	Given an MN PDA,  we can achieve the secretive rate-memory pair that matches the general achievable scheme proposed in \cite{PrivateCc} that is:
	\begin{align}
	   (M,R^s)=\Big(\frac{Nt}{K-t}+1,\frac{K}{t+1}\Big) 
	\end{align}
	where, $t=\frac{K(M-1)}{N+M-1}\in\{0,1,...,K-2\}$. 
\end{cor} 
\begin{IEEEproof} Since for the MN PDA, we have $F=\binom{K}{t},Z=\binom{K-1}{t-1}$, from Corollary \ref{cor1}, we get
	\begin{align}
	M=\frac{N\binom{K-1}{t-1}}{\binom{K}{t}-\binom{K-1}{t-1}}+1=\frac{Nt}{K-t}+1.
	\end{align}
	Substituting $g=t+1$ in \eqref{regpda_rate}, we have $R^s(M)=\frac{K}{t+1}$.
\end{IEEEproof}
\section{Proposed Scheme}\label{sec5}
In this section, we describe our proposed scheme based on a $(K,F,Z,S)$ PDA.
\subsection{Coded Caching procedure}\label{sec5A}
The main idea of the scheme is to encode each file in the library using an $(Z,F)$ non-perfect secret sharing scheme, and populate the caches with $Z$ shares of all files, and $F-Z$ randomly selected keys. After the file requests are conveyed by the users, the server broadcasts $S$ multicast messages, each of which is encrypted (XOR'ed) with a key. While the secret-sharing encoding prevents information leakage from the cache placement phase, XOR'ing the transmitted messages with keys in the delivery phase ensures that, \emph{no share} of a file that a user did not request gets leaked to the user. Hence, these keys too need to be privately placed at the caches. The steps involved in the file encoding, placement phase and delivery phase are:
\subsubsection{File Encoding}  For each file $W^n,n\in[N]$, the server first splits $W^n$ into $F-Z$ non-overlapping sub-files of equal length, $F_s:=\frac{B}{F-Z}$ bits. This forms the \emph{secret vector}, which is a $(F-Z)\times 1$ column vector, ${W}^n:=[W^n_1,W^n_2,\ldots,W^n_{F-Z}]^T$, each element of which belongs to $\mathbb{F}_{2^{F_s}}$. The server also selects $Z$ random variables uniformly and independently from the finite field $\mathbb{F}_{2^{F_s}}$ to form the \emph{encryption key vector} ${V}^n:=[V^n_1,V^n_2,\ldots,V^n_Z]^T$ of dimension $Z\times 1$. 
\par Let the \emph{share vector}, corresponding to $W^n$ be ${S}^n=[S^n_1, S^n_2,\ldots, S^n_F]^T$, where $ S^n_j\in\mathbb{F}_{2^{F_s}} \ \forall j\in[F]$ . Note that $S^n$ is a $F\times 1$ column vector $\forall n\in [N]$. Define the linear mapping $\Pi$ as
	\begin{align}\label{pi}
	\Pi:\mathbb{F}_{2^{F_s}}^{F-Z}\times \mathbb{F}_{2^{F_s}}^{Z}\mapsto \mathbb{F}_{2^{F_s}}^{F}
	\end{align}
	such that ${S}^n=\Pi({W}^n,{V}^n)$  satisfies the following conditions:
	\begin{enumerate}[(i)]
		\item	$H({W}^n|{S}^n)= 0$ [Correctness].
		\item	$H({W}^n|{S}^n_{\mathcal{Z}})= H({W}^n), \mathcal{Z}\subset[F]; |\mathcal{Z}|=Z$ [Secrecy].
	\end{enumerate}
	Condition (i) implies that all the $F$ shares of a file are sufficient to recover the file, i.e., the secret vector. Condition (ii) means that any subset of $Z$ or less shares do not reveal any information about the file. Thus, the tuple $(\Pi,W^n,S^n)$ is an $(m=Z,n=F)$ non-perfect secret sharing scheme. Each share has size $F_s=\frac{B}{F-Z}$ bits. Corresponding to each file in the library, there are $F$ shares, hence a total of $NF$ shares are generated from the library. To implement the linear mapping as in \eqref{pi}, we use an $F\times F$ Cauchy matrix $\mathbf{G}$, in finite field $\mathbb{F}_{2^r}$, where $r\geq1+\log_2{F}$, similar to the scheme in \cite{SecretiveCollud}. The formal definition of a Cauchy matrix is given in Definition \ref{cauchy}.

 For each $n\in[N]$, we concatenate the encryption key vector $V^n$ below  $W^n$ to get the vector $Y^n=[W^n;V^n]$ of dimension $F\times 1$.
	Then, the Cauchy matrix $\mathbf{G}$ is multiplied with $Y^n$ (over $\mathbb{F}_{2^r}$) to get the share vector $S^n$, that is:
	\begin{align}
	S^n_{F\times1}=\mathbf{G}_{F\times F} . Y^n_{F\times 1}    
	\end{align}
Therefore, for each $i\in[F], \forall n\in [N]$, a share $S^n_i$,
\begin{align}\label{file_enc}
S^n_i=\sum_{j=1}^Fg_{i,j}Y^n_j=\sum_{j=1}^{F-Z}g_{i,j}W^n_j+ \sum_{j=F-Z+1}^Fg_{i,j}V^n_{j-(F-Z)}.
\end{align}
Each share is a linear combination of the secret vector and encryption key vector with coefficients from Cauchy matrix.
\begin{rem}[Note on subpacketization]
	The number of sub-files which a file is split into decreases from $F$ in the non-secretive scheme \cite{PDAmain} to $F-Z$ in our secretive scheme. We see that the subpacketization level drops by an additive gap in secretive coded caching scheme, compared to the non-secretive counterpart.  This is because, to implement the $(Z,F)$ secret-sharing scheme, a file is partitioned into $F-Z$ equal-sized sub-files producing the $F-Z$-length secret vector.
\end{rem}
\subsubsection{Placement Phase}
\label{sec3A} 
In this phase, the server places shares of all files, according to \eqref{plcmnt}. Apart from shares, the server also generates $S$ random variables of size $F_s$ bits each, called \emph{unique keys}  uniformly and independently from $\mathbb{F}_{2^{F_s}}$, independent of the file distribution. They are denoted by $T_{1}, T_{2},\ldots, T_S$. The caches are filled with the shares and unique keys as follows:
\begin{itemize}
	\item User $k$ caches the $j^{th}$ share of all the files if in the PDA $\mathbf{P}$, $p_{j,k}$ is the symbol `` $*$ ''.
	\item Unique key, $T_s$ is placed in user $k$'s cache, if $s$ appears in column $k$ of $\mathbf{P}$ , i.e., if  $p_{j,k}=s, \ \forall j\in[F]$. 
\end{itemize}
Since $F-Z$ entries of each column are integers, $F-Z$ unique keys are placed in each cache. The subscript $s$ indicates the time slot in which the respective key is used to encrypt the message during the delivery phase. The cache content, $Z_k$ of user $k$ is precisely:
\begin{align*}
Z_k=\big\{\cup_{j\in[F]:p_{j,k}=*}S^n_j,\hspace{0.1 in}\forall n\in [N]\big\}\bigcup\big\{\cup_{j\in[F]:p_{j,k}=s}T_s\big\}.
\end{align*}
The portion of cache memory occupied by the shares is $NZF_s=\frac{NZ}{F-Z}B$ bits while that occupied by unique keys is $(F-Z)F_s=B$ bits. Normalizing by file size, the cache memory at every user denoted by $M$ is $\frac{NZ}{F-Z}+1$, which always exceeds $1$.

\subsubsection{Delivery Phase}
In the content delivery phase, the indices of files demanded by the users, represented by the demand vector $\mathbf{d}=(d_1, d_2,\ldots,d_K)$ are revealed. We focus on the case when $N\geq K$. The server sends XOR'ed combinations of shares and unique keys over the shared link. Delivery takes place in $S$ time slots, where each transmission corresponds to an integer $s\in [S]$. We denote the transmission occurring in time slot $s$ as $X_{\mathbf{d},s}$. The overall transmitted messages is given by the union of them, i.e., $X_{\mathbf{d}}=\bigcup_{s\in [S]} X_{\mathbf{d},s}$, where
\begin{align}\label{transmit}
X_{\mathbf{d},s}=\Big(\bigoplus_{p_{j,k}=s, j\in[F], k\in[K]}S^{d_k}_j\Big)\oplus T_s.   
\end{align}
\begin{rem}
	Since each message is XOR'ed with a unique key $T_s$, the delivery is also secure against external eavesdroppers who might wiretap on the shared link. This means, the eavesdroppers do not obtain any information about the shares, (hence the files), because they do not have access to the uniformly distributed keys $T_s, s\in [S]$. 
\end{rem}
\subsection{Rate calculation:}\label{pda_rate}
Corresponding to each integer $s$ in the PDA, the message $X_{\mathbf{d},s}$ as in \eqref{transmit} is transmitted. In total, there are $S$ transmissions, each of size $F_s=\frac{B}{F-Z}$ bits leading to a transmission of $\frac{SB}{F-Z}$ bits over the shared link. The delivery rate normalized by file size is therefore $\frac{S}{F-Z}$ as stated in Theorem \ref{scc_rate}.

\subsection{Proof of correctness} 
We show that, each user can obtain the file it wants, by
utilizing the transmissions and its cache contents. From the placement phase, $Z$ out of $F$ shares of all the files are available to every user. Hence, $Z$ shares of the requested file are already present in the cache of a given user. To completely recover the secret, the remaining $F-Z$ shares of the requested file need
to be transmitted in the delivery phase. For user $k$ requesting $W^{d_k}$, out of the $S$ transmissions made by the server in \eqref{transmit}, only
those made in time-slots corresponding to the distinct integers present in the $k^{th}$ column of PDA contain the missing shares of $W^{d_k}$. Using  the $F-Z$ unique keys present in its cache, it recovers the XOR’ed combination
of shares. In each such combination, except for the missing
share of the requested file, the remaining shares are already
cached by the user (by C3. of Definition \ref{defPDA}). Therefore, from
each such message, the user obtains a distinct share of $W^{d_k}$ .
Hence, the $F-Z$ shares of $W^{d_k}$ not cached by the user are
obtained. This way, the share vector $S^{d_k}$ is retrieved $\forall k\in [K]$
\par Once the share vector is available, user $k$ needs to evaluate $\mathbf{G}^{-1}S^{d_k}$. Note that $\mathbf{G}$ is known to the users. Since a Cauchy matrix is full rank, $\mathbf{G}^{-1}$ always exists. Thus, user $k$ can evaluate $Y^{d_k}$, the first $F-Z$ elements of which yields $W^{d_k}$. Hence, \eqref{corr} is satisfied.
\subsection{Proof of secrecy}  The proof is similar to that in \cite{SecretiveCollud}. Since every file in the library follows the same caching algorithm, the users must not gain any information about the content of \emph{any file} in their caches. First, we show that, user $k$ can obtain no information about $W^{d_k},\forall k\in[K]$ from its cache contents. Let an arbitrary $Z$-sized subset of $S^{d_k}$ be denoted by $S^{d_k}_{i_1},S^{d_k}_{i_2},\ldots,S^{d_k}_{i_Z}$. Therefore, with the file encoding in \eqref{file_enc},
	\begin{align}
	\begin{bmatrix}
	S^{d_k}_{i_1}\\S^{d_k}_{i_2}\\ \vdots\\S^{d_k}_{i_Z}
	\end{bmatrix}
	&=
	\begin{bmatrix}
	g_{i_1,1}&g_{i_1,2}&\ldots&g_{i_1,F}\\
	g_{i_2,1}&g_{i_2,2}&\ldots&g_{i_2,F}\\
	\vdots& \vdots & \ddots & \vdots\\
	g_{i_Z,1}&g_{i_Z,2}&\ldots&g_{i_Z,F}\\
	\end{bmatrix}
	\begin{bmatrix}
	W^{d_k}_1\\
	W^{d_k}_2\\
	\vdots\\
	W^{d_k}_{F-Z}\\
	V^{d_k}_1\\
	V^{d_k}_2\\
	\vdots\\
	V^{d_k}_Z\\
	\end{bmatrix}
	\end{align}
	\begin{align}
	&=\mathbf{G_1}
	\begin{bmatrix}
	W^{d_k}_1\\
	\vdots\\
	W^{d_k}_{F-Z}\\
	\end{bmatrix}
	+ \mathbf{G_2}
	\begin{bmatrix}
	V^{d_k}_1\\
	\vdots\\
	V^{d_k}_Z\\
	\end{bmatrix}
	=\mathbf{G_1}W^{d_k}+\mathbf{G_2}V^{d_k}
	\end{align}
	where $\mathbf{G_1}$ and $\mathbf{G_2}$ are submatrices of $\mathbf{G}$ of dimensions $Z\times(F-Z)$ and $Z\times Z$ respectively.
	\par For the subset of shares to leak information, the encryption key vector $V^{d_k}$ must be decoupled from the corresponding secret vector $W^{d_k}$, i.e., $\mathbf{G_1}W^{d_k}\neq\mathbf{0}$ and $\mathbf{G_2}V^{d_k}=\mathbf{0}$. By Definition \ref{cauchy}, since all submatrices of a Cauchy matrix are full rank, the columns of $\mathbf{G_2}$ are linearly independent. Hence, such $V^{d_k}$ does not exist. This implies, that a linear combination involving only the sub-files $W^{d_k}$ can not be obtained. Hence information leakage from caching is 0.
\par Now, to show that an arbitrary user $k$ cannot obtain any information about $W^{[N]\setminus d_k}$ from the transmitted messages, observe that each message is encrypted with a key which is available only to those users for whom the message is useful. A message is useful to a user if the user can retrieve a missing share from it. For a transmission, $X_{\mathbf{d},s}$ where $s$ does not appear in the $k^{th}$ column of the PDA, user $k$ has no knowledge of $T_s$. Hence, user $k$ can obtain no information about the linear combination of shares encrypted by this key. Therefore, \eqref{sec} is satisfied.

\begin{exmp}
	Consider the $3-(6,4,2,4)$ PDA $\mathbf{P}$ as given in Example \ref{ex1}.
	With secrecy, this corresponds to a $(K,M,N)$ caching system with $N=K=6$ and $M=\frac{NZ}{F-Z}+1=7$ units. The files, $W^1,W^2,\ldots,W^6$ of $B=6$ bits each are split into $F-Z=2$ sub-files of length $B/2=3$ bits. Each sub-file is an element of $\mathbb{F}_{2^3}$. For every $n\in [6]$, the sub-files of $W^n$ are $W^n_1$ and $W^2_2$. They form the secret vector $W^n=[W^n_1 \ , W^n_2]^T$. Further, $\forall n \in [6]$ the server chooses $V^n_1$ and $V^n_2$ independently from $\mathbb{F}_{2^3}$ and let $V^n=[V^n_1 \ ,  V^n_2]^T$. We premultiply each $Y^n=[W^n; V^n]$ with a $4\times 4$- Cauchy matrix over  $\mathbb{F}_{2^3}$. Let $\alpha$ be the primitive element of $\mathbb{F}_{2^3}$ satisfying $\alpha^3+\alpha+1=0$. Let $X=\{0,2,4,6\}, Y=\{1,3,5,7\}$. Then, we obtain the following Cauchy matrix:
	\begin{align*}
	\mathbf{G}=\hspace{-1mm}
	\begin{bmatrix}
	1 &  \alpha^2+\alpha & \alpha & \alpha^2\\
	\alpha^2+\alpha & 1 & \alpha^2 & \alpha\\
	\alpha & \alpha^2 & 1 & \alpha^2+\alpha\\
	\alpha^2 & \alpha & \alpha^2+\alpha & 1
	\end{bmatrix}
	\hspace{-1mm}=
	\begin{bmatrix}
	1 &  6 & 2 & 4\\
	6 & 1 & 4 & 2\\
	2 & 4 & 1 & 6\\
	4 & 2 & 6 & 1
	\end{bmatrix}
	\end{align*}
	The four shares corresponding to $W^n,\forall n\in[6]$ are given by:
	\begin{align}
	\begin{bmatrix}
	S^n_1\\S^n_2\\S^n_3\\S^n_4
	\end{bmatrix}
	&=
	\begin{bmatrix}
	1 &  6 & 2 & 4\\
	6 & 1 & 4 & 2\\
	2 & 4 & 1 & 6\\
	4 & 2 & 6 & 1
	\end{bmatrix}
	\begin{bmatrix}
	W^n_1\\W^n_2\\V^n_1\\V^n_2
	\end{bmatrix}.
	\end{align}
	Apart from shares, the server generates the unique keys $T_1,T_2,T_3,T_4\in \mathbb{F}_{2^3}$. The cache content $\forall n\in [6]$ is:
		\begin{align*}
		Z_1 & =S^n_1, S^n_3, T_1, T_3; \ 
		Z_2=S^n_2, S^n_4, T_2, T_4 ;\\ 
		Z_3 &=S^n_1, S^n_2, T_1, T_2; \
		Z_4 =S^n_3, S^n_4, T_3, T_4 ; \\
		Z_5 &=S^n_1, S^n_4, T_2, T_3; \
		Z_6 =S^n_2, S^n_3, T_1, T_4.
		\end{align*}
	Without loss of generality, consider $\mathbf{d}=(1,2,3,4,5,6)$. The server sends the following messages in the delivery phase:
	\begin{align*}
	\mbox{Time slot 1}: S^1_2\oplus S^3_3 \oplus S^6_1\oplus T_1\\
	\mbox{Time slot 2}: S^2_1\oplus S^3_4 \oplus S^5_2\oplus T_2\\
	\mbox{Time slot 3}: S^1_4\oplus S^4_1 \oplus S^5_3\oplus T_3\\
	\mbox{Time slot 4}: S^2_3\oplus S^4_2 \oplus S^6_4\oplus T_4
	\end{align*}
Using these transmissions and the cache contents, every user can retrieve all the $4$ shares of the file it wants. For example, user 1 wants all the shares of $W^1$. It gets $S^1_1$ and $S^1_3$ from its cache, and needs $S^1_2$ and $S^1_4$ from the server transmissions. Since it has $T_1$ and $T_3$, it obtains the linear combinations $S^1_2\oplus S^3_3 \oplus S^6_1 \text{ and } S^1_4\oplus S^4_1 \oplus S^5_3$ by XOR'ing out $T_1$ and $T_3$ respectively. Now, user 1 has the $1^{st}$ and $3^{rd}$ shares of all files in its cache, and can eliminate the interfering shares $S^3_3 \oplus S^6_1$ and $S^4_1 \oplus S^5_3$ to obtain $S^1_2$ and $S^1_4$ respectively. From $S^1_1,S^1_2,S^1_3$ and $S^1_4$ it can calculate $W^1_1$ and $W^1_2$ by evaluating $W^1_1=S^1_1\oplus 6S^1_2 \oplus 2S^1_3\oplus 4S^1_4$ and $W^1_2=6S^1_1\oplus S^1_2 \oplus 4S^1_3\oplus 2S^1_4$. Also, it does not obtain any additional share of $W^2,W^3$ or $W^4$. Given the access structure of the $(2,4)$ non-perfect secret-sharing scheme, having $2$ shares of these files maintains \eqref{sec}.
The  secretively achievable rate of this scheme is $R^s(7)=2$ at a subpacketization of $2$ while the scheme in \cite{PrivateCc} for the above caching parameters $(K,M,N)$ and $t=\frac{6(7-1)}{7+6-1}=3$ achieves a rate of $\frac{6}{4}=1.5$ at a subpacketization of $\binom{6-1}{3}=10$.
\end{exmp}

\begin{table*}[tp]
	\caption{Performance of secretive coded caching scheme from PDAs with respect to the general achievability scheme in \cite{PrivateCc}}
	\begin{center}
	\begin{tabular}{|c|c|c|c|c|c|c|}
		\hline
		\multirow{2}{*}{PDA} & \multirow{2}{*}{$K$} & \multirow{2}{*}{$M$} & \multicolumn{2}{c|}{Rate} & \multicolumn{2}{c|}{Subpacketization level}\\
		\cline{4-7}
		& & & $R^s_{MN}$\cite{PrivateCc} & $R^s_{PDA}$ & $F^s_{MN}$\cite{PrivateCc} & $F^s_{PDA}$\\
		\hline
		\hline
		\thead{\cite{PDAmain},Theorem 4,\\$m=1$\\$2-(2q,q,1,q^2-q)$} & $2q$ & $1+\frac{N}{q-1}$ & $\frac{2}{3}q$ & $q$ &  $(q-1)(2q-1)$ & $q$ \\
		\hline
		\thead{\cite{PDAmain},Theorem 5,\\$m=1$\\$2-(2q,q^2-q,(q-1)^2,q)$} & $2q$ & $1+N(q-1)$ & $\frac{q}{(q-1)+\frac{1}{2}}$  & $\frac{q}{q-1}$  & $2q-1$ & $q$ \\
		\hline
		\thead{\cite{PDAhyperg},Theorem 14,\\ $a=1,b=2,n\geq 3$\\$3-\bigg(\binom{n}{2},n,2,\binom{n}{3}\bigg)$} & $\frac{n(n-1)}{2}$ & $1+N\frac{2}{n-2}$  &$\frac{n-1}{2}$ & $\frac{n(n-1)}{6}$ & $\approx 2^{\frac{(n+1)(n-2)}{2} h\big(\frac{2(n-1)}{(n+1)(n-2)}\big)}*$  & $n-2$\\
		\hline
		\thead{\cite{PDAhyperg},Theorem 18,\\$m=2,t=1$\\$2-(2q,q^2,q,q^3-q^2)$} & $2q$ & $1+\frac{N}{q-1}$  & $q\frac{q}{2(q+1)}$ & $q$ &  $(q-1)(2q-1)$ & $q(q-1)$\\
		\hline
	\end{tabular}
	\end{center}
	\label{comp}
	 $*h(t)$ is the binary entropy function.
\end{table*}
\begin{table}[htbp]
    \centering
        \caption{Comparison of caching scheme with and without secrecy for $(m+1)$ -$(q(m+1),q^m,q^{m-1},q^{m+1}-q^m)$ PDA \cite{PDAmain}}
    \begin{tabular}{|c|c|c|}
    \hline
      & Without secrecy\cite{PDAmain}  &  With secrecy\\
      \hline
    Number of users & $q(m+1)$ & $q(m+1)$\\
    \hline
    Memory & $\frac{N}{q}$ & $1+\frac{N}{q-1}$\\
    \hline
    Subpacketization level & $q^m$ & $q^m-q^{m-1}$\\
    \hline
    Rate & $q-1$ & $q$\\
    \hline
    \end{tabular}
\label{tab2}
\end{table}
\section{Performance Comparison}
\label{sec6}
Table \ref{comp} shows the performance of our scheme applied on several PDAs \cite{PDAmain,PDAhyperg} and that of the scheme in \cite{PrivateCc} in terms of secretively achievable rate and subpacketization level. We refer to the rate and subpacketization levels corresponding to the scheme in \cite{PrivateCc} as $R^s_{MN}$ and $F^s_{MN}$ respectively, where $R^s_{MN}$ is the rate in \eqref{eq1} and $F^s_{MN}=\binom{K-1}{t},t=\frac{K(M-1)}{N+M-1}$ and the rate and subpacketization levels corresponding to the PDA based scheme as $R^s_{PDA}$ and $F^s_{PDA}$ respectively. Every PDA in Table \ref{comp} yields a $(K,M,N)$ secretive coded caching system, and for each of them, $F^s_{PDA}<F^s_{MN}$ at the expense of slight increase in secretively achievable rate $R^s_{PDA}$ compared to $R^s_{MN}$. In Table \ref{tab2}, our scheme is compared against the non-secretive scheme in \cite{PDAmain} based on the $(m+1)$ regular $(q(m+1),q^m,q^{m-1},q^{m+1}-q^m)$ PDA constructed in \cite{PDAmain}. On imposing the secrecy constraint, the delivery rate is higher by a unit, at the expense of a larger memory, however providing an advantage of $q^{m-1}$ units in terms of subpacketization.
\section{Conclusion}\label{sec7}
We presented a feasible secretive coded caching scheme based on PDAs which, due to the structure of PDA is characterised by sub-exponential level of subpacketization. The general achievable scheme in \cite{PrivateCc} arises from the MN PDA and thereby gets subsumed by the proposed scheme. As a consequence of the non-perfect secret sharing scheme used to encode each file, the subpacketization level for any secretive coded caching scheme is reduced by an additive gap.
\section*{Acknowledgment}
This work was supported partly by the Science and Engineering Research Board (SERB) of Department of Science and Technology (DST), Government of India, through J. C. Bose National Fellowship to B. Sundar Rajan.

\end{document}